# Field-induced topological phase transition from three-dimensional Weyl semimetal to two-dimensional massive Dirac metal in $ZrTe_5$

Guolin Zheng[1*], Xiangde Zhu[1*], Yequn Liu[2], Jianwei Lu[1,3], Wei Ning[1§)], Hongwei Zhang[1,3], Wenshuai Gao[1,3], Yuyan Han[1], Jiyong Yang[1], Haifeng Du[1,4], Kun Yang[5§)], Yuheng Zhang[1,3,6] and Mingliang Tian[1,4,6§)]

[1]*Anhui Province Key Laboratory of Condensed Matter Physics at Extreme Conditions, High Magnetic Field Laboratory of the Chinese Academy of Science, Hefei 230031, P. R. China.*

[2]*Institute of Coal Chemistry, Chinese Academy of Sciences, Taiyuan 030001, P. R. China; Analytical Instrumentation Center, Institute of Coal Chemistry, Chinese Academy of Sciences, Taiyuan 030001, P. R. China.*

[3]*Department of physics, University of Science and Technology of China, Hefei 230026, P. R. China.*

[4]*Department of Physics, College of Physics and Materials Science, Anhui University, Hefei 230601, P. R. China.*

[5]*National High Magnetic Field Laboratory, Florida State University, Tallahassee, Florida 32306-4005, USA.*

[6]*Collaborative Innovation Center of Advanced Microstructures, Nanjing University, Nanjing 210093, P. R. China.*

[*] Those authors contribute equally to this work.

[§]To whom correspondence should be addressed. E-mail: kunyang@magnet.fsu.edu (K. Y.), ningwei@hmfl.ac.cn (W. N.),tianml@hmfl.ac.cn (M. T.).

# Abstract

Symmetry protected Dirac semimetals can be transformed into Weyl semimetals by breaking the protecting symmetry, leading to many exotic quantum phenomena such as chiral anomaly and anomalous Hall effect. Here we show that, due to the large Zeeman g factor and small band width along *b*-axis in Dirac semimetal $ZrTe_5$, a magnetic field of about 8 T along *b*-axis direction may annihilate the Weyl points and open up a two-dimensional (2D) Dirac mass gap, when the Zeeman splitting exceeds the band width along *b*-axis. This is manifested by a sharp drop of magnetoresistance (MR) above 8 T, which is probably due to additional carriers induced by the orbital splitting of the zeroth Landau level associated with the 2D Dirac point, which is descendant of the original Weyl points. Further evidence of the additional carriers is provided by the Hall effect and different anisotropic magnetoresistance (AMR) in low and high field regions. Our experiment reveals a probable topological quantum phase transition of field induced Weyl points annihilation in Dirac semimetal $ZrTe_5$ and gives an alternative explanation for the drop of MR at high field.

Three-dimensional (3D) Dirac and Weyl semimetals have attracted extensive attentions recently, due to their exotic transport properties and potential impacts on information technology [1-13]. Dirac semimetal, holding a pair of Weyl points coinciding at the Dirac point, is usually unstable unless protected by some special symmetries such as time reversal symmetry (TRS), spatial inversion symmetry as well as crystalline point-group symmetry [14-17]. Thus a magnetic field B, which breaks time reversal symmetry, can turn a Dirac semimetal to a Weyl semimetal, which is more robust since the Weyl points can only disappear by annihilation of pairs with opposite topological charges [18]. The recently discovered Dirac semimetal phase in layered material $ZrTe_5$ provides an ideal platform to explore the exotic quantum phenomena in TRS-broken Dirac semimetals, since in this layered material a modest magnetic field can drive the system into the quantum limit [19, 20]. In addition, compared with other topological semimetals, layered material $ZrTe_5$ has very weak interlayer coupling [21] as well as large Zeeman g factor according to recent magneto infrared spectroscopy study [19]. A small band width in the interlayer direction can be anticipated due to the weak interlayer coupling, which allows for a probable interplay with the Zeeman splitting when a sufficiently strong magnetic field is applied along the interlayer direction. Some unusual quantum phenomena can be anticipated due to the large Zeeman splitting, especially when it exceeds the band width in the interlayer direction, a regime never explored experimentally before.

In this paper, we show that in the interlayer direction of $ZrTe_5$ bulk crystals, Zeeman splitting at high field will likely exceed the band width and quench dispersion in this direction, and turn the 3D (massless) Weyl points into a *massive* 2D Dirac point by annihilating the Weyl points when they reach Brillouin zone (BZ) boundary. Such a topological quantum phase transition is revealed by a sharp drop of MR at high field，which is probably due to the orbital splitting of the massive 2D Dirac point when the system is approaching to the zeroth Landau Level (LL). Our model of field-induced Weyl points annihilation in Dirac semimetal $ZrTe_5$ provides an alternative explanation for the drop of MR at high field.

$ZrTe_5$ single crystal has an orthorhombic layered structure with space group *Cmcm* ($D_{2h}^{17}$) [21]. Despite its relatively low-symmetry crystal structure, this layered material has been demonstrated to be a 3D Dirac semimetal recently by magneto-infrared spectroscopy study [19, 22], ARPES [2] and electrical transport experiments [2, 20]. In $ZrTe_5$, Zr atoms lie at the centers of right trigonal prisms of Te atoms stacked along the $a$-axis, while the parallel chains of trigonal prisms are coupled in the c-axis direction via zigzag chains of Te atoms to form two-dimensional *a-c* plane, then stacked along the $b$-axis into a crystal [23]. The detailed crystal growth and characterization are presented in Supplementary information (Part I). We performed transport measurements in different $ZrTe_5$ bulk crystals (sample ♯1~sample ♯4) with current along $a$-axis, which is the most conductive direction.

Fig. 1(a) shows the magnetoresistance (MR) of sample ♯1 at various temperatures with magnetic field applied along $b$-axis. In low temperature region, the MR exhibits strong Shubnikov–de Hass (SdH) oscillations and the MR reaches a maximum at a critical field $B^*$ of about 8 T at 2 K, followed by a sharp drop. At the first glance of the MR curve at 2 K, it seems that the MR peak near 8 T is due to the quantum oscillation or the Landau level (LL) splitting. However, with increasing the temperature, the peak position of MR shifts toward the high field region and disappears at T>70 K in our measured range. Such a MR peak cannot be attributed to LL splitting or the quantum oscillation. This is because temperature can only smear the splitting and weaken the amplitudes of the SdH oscillations *without* shifting the peak position. To give a clear demonstration of the Landau level, we plot $dR/dB$ versus 1/B under different temperatures in Fig. 1(b). It clearly shows that a moderate magnetic field of about 5 T can drive the system into the first LL $n$=1, and the peak positions of low Landau index ($n \leq 2$) are split into several sub-peaks at 2 K. Such split behavior of the LL can be attributed to the removal of spin degeneracy due to the spin Zeeman splitting [6, 19].

To have a better understanding of the SdH oscillation, we have tracked the angular-dependent magnetoresistance (ADMR) of sample #1 at 2 K under different tilted angle with B rotating in both *b-a* and *b-c* planes, as shown in Fig. 2(a) and 2(c), respectively. The tilted angle $\theta$ ($\varphi$) is defined between B and *b*-axis rotating in the *b-a* (*b-c*) planes, respectively. Fig. 2(b) and 2(d) shows the dR/dB as the function of 1/Bcos(θ) and 1/Bcos(φ), respectively. As we can see, the SdH oscillations in the low field region reveal a 3D Fermi surface, as marked by the dark dashed arrows. While in high field region, the LLs split into several sub-peaks due to the spin Zeeman splitting. Note that there is a remarkable shoulder near B= 3 T which cannot be attributed to the spin Zeeman splitting, which we do not have a simple explanation at the moment.

Generally, the sub-peaks of spin Zeeman splitting should superimpose on the oscillation peaks of 3D bulk background, which should also exhibit a 3D tendency. While in our observations, the 3D nature of bulk background is suppressed gradually as field increases, as indicated by the blue dashed arrows in Fig. 2(b) and 2(d), respectively. Above 8 T, the sub-peaks exhibit an angular-dependent tendency, as marked by the red dashed arrows, indicating a vanishing 3D bulk background and a quasi-2D nature of electron motion at high field.

The Zeeman energy can be described as $\varepsilon = g\mu_B B$, where $\mu_B$ is the Bohr magneton. The g factor can be extracted by $g = 2F\Delta(1/B)\, m_0/m^*$, where $\Delta(1/B)$ is the splitting spacing of the oscillation peaks and $F$ is the oscillation frequency [24]. According to our recent studies [20], The effective cyclotron mass has $m^* \sim 0.026\, m_0$ for *a*-*c* plane carriers in $ZrTe_5$. The calculated the effective g factor is about 32, which is comparable with recent experiments [19, 25]. Such a large effective Zeeman g factor hints that the Zeeman splitting may compete with the band width in interlayer direction at high field. To reveal the potential interplay between Zeeman splitting and band width along *b*-axis direction, we have systematically analyzed the band width via the low field SdH oscillations (Supplementary Part II). The calculated band width in our $ZrTe_5$ single crystals along *b-axis* is about 10 *meV* to 19 *meV*. At 8 T, the Zeeman splitting energy $g\mu_B B/2$ is about 8 *meV*, which is comparable with the band width along *b*-axis direction. Further increasing the field above 8 T, the Zeeman splitting may exceed the band width along *b*-axis and quench the dispersion along this direction, driving the 3D Fermi surface to a quasi-2D Fermi surface, as we demonstrate below.

To gain further insight into the quasi-2D character in high field region, we consider the low-energy effective Hamiltonian of a single Weyl point. In the absence of the magnetic field:

$$H = \sum_{j=1}^{3} v_j \sigma_j p_j . \quad (1)$$

Here $p$ is momentum, $\sigma$ are Pauli matrices corresponding to electron spin. $v_{j=1,2,3}$ are the components of Fermi velocity along the three axes. The Weyl point is chosen to be at $k$=0. Applying a magnetic field B along $\hat{z}$, due to the Zeeman effect, the 3rd term in Eq. (1) becomes:

$$H_z = (v_3 p_3 - \Delta_z)\sigma_3, \quad (2)$$

where $\Delta_z = g\mu_B B/2$ is the Zeeman splitting. For small B and $\Delta_z$, it shifts the Weyl point to $k = (0,\ 0,\ \Delta_z/v_3)$. However, when B reaches certain critical value, the Weyl point reaches the BZ boundary and cannot move further. Beyond this point the coefficient of $\sigma_3$ in Eq. (2) will never be zero, thus the electron energy will never be zero, and the 3D Weyl point is lost. The $\sigma_3$ term in Eq. (2), whose coefficient is always non-zero, plays a role very similar to a Dirac mass for a 2D Dirac Hamiltonian.

Another way to understand this is the pair of Weyl points, initially split by the Zeeman field moving apart from each other (say from the BZ center), start to move toward each other as the Zeeman splitting becomes sufficiently strong due to the periodicity of the BZ, and eventually meet and annihilate each other at the BZ boundary, as illustrated in Fig. 3(a)-3(c). As a result a mass gap is opened up in the electronic spectrum and the system is transformed from the 3D massless Weyl semimetal to a quasi-2D massive metal. Such field induced Weyl points annihilation is a topological quantum phase transition and has never been reported experimentally before.

When the two Weyl points meet at the Brillouin zone boundary, the two Weyl points now need to be taken into account on equal footing. For simplicity we assume in the following that the velocity is isotropic and the annihilation of two Weyl points happens at k=0, in which case the low-energy effective Hamiltonian near the annihilation transition is

$$H = v\left(\sigma_x p_x + \sigma_y p_y\right) - \sigma_z[\, \alpha k_z^2 + \beta(B - B^*)]. \qquad (3)$$

Here $\alpha$ and $\beta$ are constants assumed to be positive. For $B < B^*$ there are two Weyl points at $k_z = \pm\sqrt{\beta(B^* - B)/\alpha}$, which merge at $B = B^*$. For $B > B^*$ the coefficient of $\sigma_z$ never vanishes and there is always a gap in the spectrum.

Now assuming the Zeeman splitting is large enough such that the dispersion along $\hat{z}$ direction can be neglected so the system is effectively 2D-like, we have

$$H = v\left(\sigma_x p_x + \sigma_y p_y\right) - \Delta_Z \sigma_z\,, \qquad (4)$$

which is the 2D massive Dirac Hamiltonian.

We now consider the orbital effect that couples to the motion in the *a-c* plane when the system's Fermi energy approaches the zeroth LL. In the presence of magnetic field we need to perform the standard minimum substitution of

$$\vec{p} \Longrightarrow \vec{\Pi} = \vec{p} + e\vec{A}(\vec{r})/c, \qquad (5)$$

where $\vec{\Pi}$ is the mechanical momentum.

We have

$$H = \begin{pmatrix} -\Delta_Z & v(\Pi_x - i\Pi_y) \\ v(\Pi_x + i\Pi_y) & \Delta_Z \end{pmatrix} = \begin{pmatrix} -\Delta_Z & \epsilon_D a^+ \\ \epsilon_D a & \Delta_Z \end{pmatrix}, \quad (6)$$

where

$$a = \frac{\ell}{\sqrt{2}\hbar}(\Pi_x + i\Pi_y), \quad a^+ = \frac{\ell}{\sqrt{2}\hbar}(\Pi_x - i\Pi_y), \quad (7)$$

in which $\ell = \sqrt{\hbar c/eB}$ is the magnetic length such that $[a, a^+] = 1$, and the parameter $\epsilon_D = \sqrt{2}\hbar v/\ell$. Now we can see that the energy of zeroth LL $|0\rangle$ has shifted from zero energy to $-\Delta_z$. The energy shift of zeroth LL $\Delta_z$ is $\beta(B - B^*)$ in high field region, where $\beta$ is approximately $g\mu_B$. The estimated $\Delta_z$ is about 7 meV near 15 T, above which $\Delta_z$ will exceed the Fermi energy $E_F \sim 7\ meV$ (Supplementary Part II) of our $ZrTe_5$ crystals and break the particle-hole symmetry in the energy spectrum, as illustrated in Fig. 3(e).

This is in sharp contrast to the case $B < B^*$, where in the Weyl semimetal phase the spectrum is strictly particle-hole symmetric, with or without the magnetic field. Thus the topological transition associated with the field-induced annihilation of the Weyl points reorganizes the electronic spectrum and induces additional carriers. This is consistent with our observation in Fig. 4, where Hall resistance of sample #2 shows an anomaly near MR peak position, hinting that extra carriers have been introduced by the orbital effect. Also, those extra carriers will enhance the conductivity and lead to a drop of MR near 8 T. This orbital splitting in 2D massive Dirac metal in the zero-mode LL is similar to the valley polarization of zeroth LL in transition metal dichalcogenides [26-31], where the valleys shift in opposite directions under magnetic field due to the different chiralities.

Further evidence for the orbital splitting in high field region comes from the anisotropic magnetoresistance (AMR) of sample #1 in a rotating magnetic field, which provides a better insight on the anisotropy of the Fermi surface, as did in some other Dirac systems [32-35]. Fig. 5 shows the AMR with B rotating *consecutively* in the *b*-*c* plane at an angle $\varphi$ to *b*-axis. In layered material $ZrTe_5$, the carrier in *a*-*c* plane has a maximal mobility and large MR, while in the other planes, the MR will be much smaller than that in *a*-*c* plane [36, 37]. Thus, as $\varphi$ consecutively increases from $0^{\circ}$ to $360^{\circ}$, the MR will change accordingly, and a 2-fold symmetry of AMR is formed in low field region (B<8 T). While above 8 T, the zeroth LL will shift away from zero energy consecutively as the increasing of the field, due to the orbital effect. Correspondingly, the magnetoresistance will be largely suppressed and the observed 2-fold symmetry exhibits a deformation. Similar behavior can also be seen for magnetic field rotating in the *b*-*a* plane under different temperatures (Supplementary Part IV). As temperature increases, the orbital splitting is smeared out gradually, as we can see in Fig. 5(b)-5(e) for T= 10 K, 30 K, 50 K and 80 K respectively. Recently, a quasi-2D Fermi surface is revealed in semimetal $ZrTe_5$ single crystals [38], and the drop of MR at high field is attributed to the formation of the density wave in quantum limit, which will generate a mass gap [39]. However, a number of recent reports [20，22，40] indicated that $ZrTe_5$ crystals exhibit a 3D Fermi surfaces. We believe that the many-body interaction is weak in our system due to the 3D Fermi surface with relative large Fermi velocity in interlayer direction. Since we have experimentally observed a clear transformation of the Fermi surface from 3D to quasi-2D, we therefore attribute such a resistance drop at quantum limit to the field-induced Weyl points annihilation. This model provides an alternative understanding on the exotic phenomena in Dirac semimetals.

In conclusion, we report a field-induced topological phase transition from 3D Weyl semimetal to 2D massive Dirac metal in $ZrTe_5$ when the Zeeman splitting exceeds the band width along the layered direction (i.e., the *b*-axis). This is manifested by a sharp drop of MR and Hall anomaly above 8 T induced by the orbital effect. Further evidence is provided by the different AMR under various rotating magnetic fields. Our experiment reveals a probable field-induced Weyl points annihilation, which can stimulate further research on field-induced topological quantum phase transition in topological semimetals.

## Acknowledgments

This work was supported by the National Key Research and Development Program of China, No. 2016YFA0401003, 2017YFA0303201, the Natural Science Foundation of China (Grant No. 11174294, 11574320, 11374302, 11204312, 11474289, U1432251), National Science Foundation of USA (Grant No. DMR-1442366 and DMR-1157490), the CAS/SAFEA international partnership program for creative research teams of China..

# Figure Captions

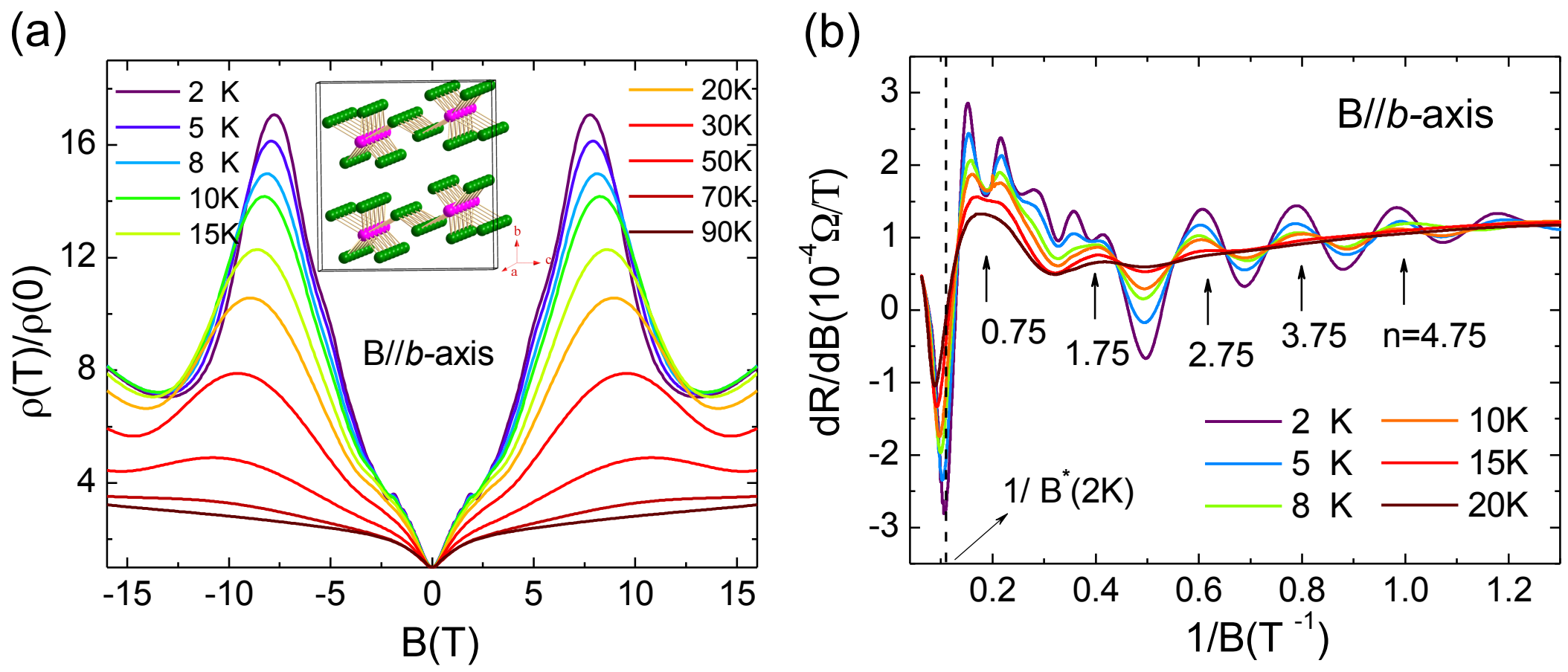

FIG. 1 (Color online). Temperature dependent magnetoresistivity and the SdH oscillations in sample #1. (a) The magnetoresistivity under different temperatures. The current is injected along *a*-axis. Inset: the crystal structure of $ZrTe_5$. (b) The SdH oscillation component as function of 1/B at different temperatures. The peak positions of dR/dB are defined as integer LL indices. Note that there is a 1/4 period shift of the derivative of resistance.

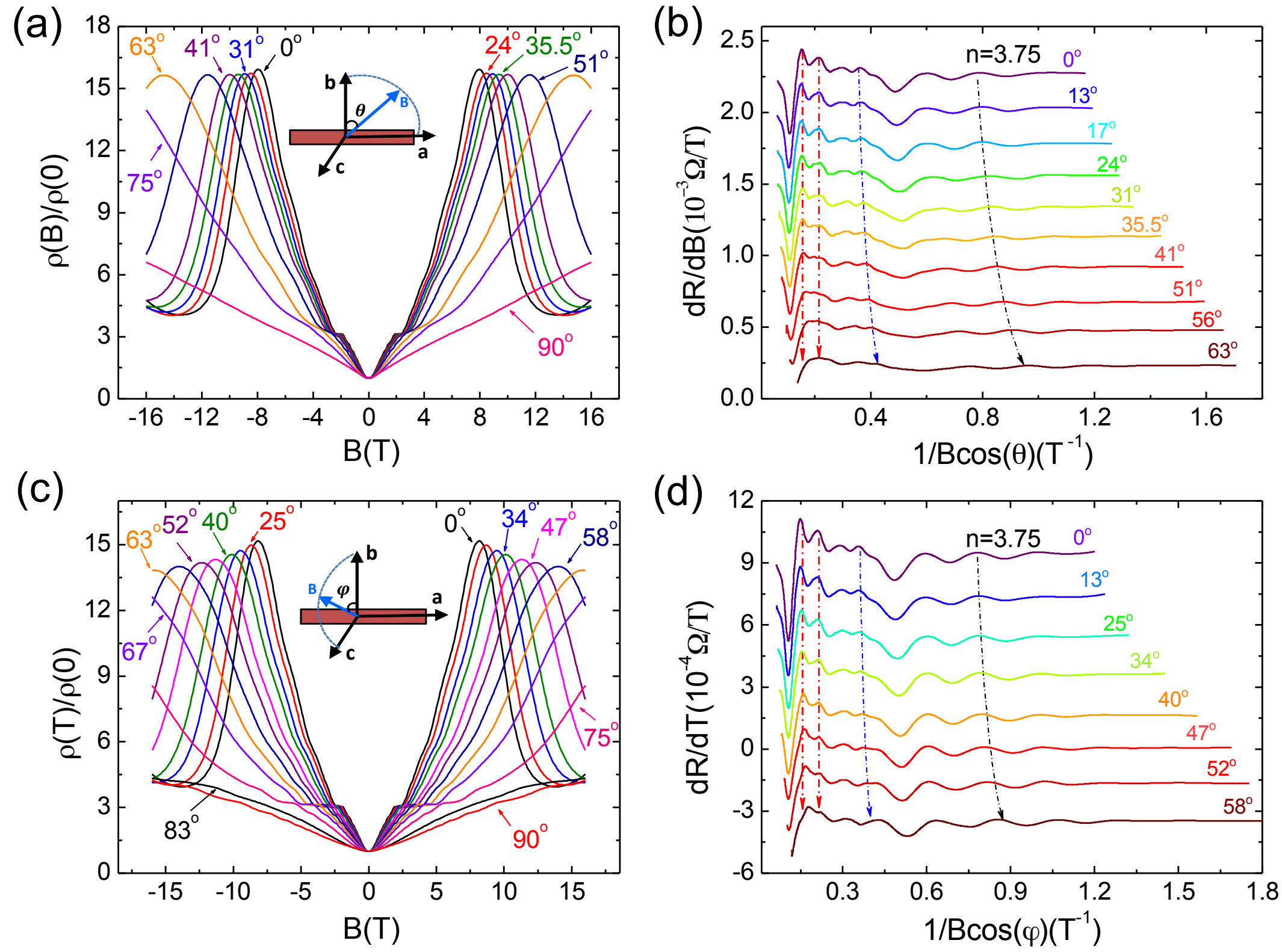

FIG. 2 (Color online). Angular dependent magnetoresistance (ADMR) in sample ♯1at 2 K. (a) The ADMR with B tilted in *b*-*a* plane under different tilted angle θ. (b) The angular dependence of the resistance oscillations versus 1/Bcos(θ). (c) and (d) are respectively the ADMR and the resistance oscillations with B tilted in *b*-*c* plane with an angle φ to *b*-axis.

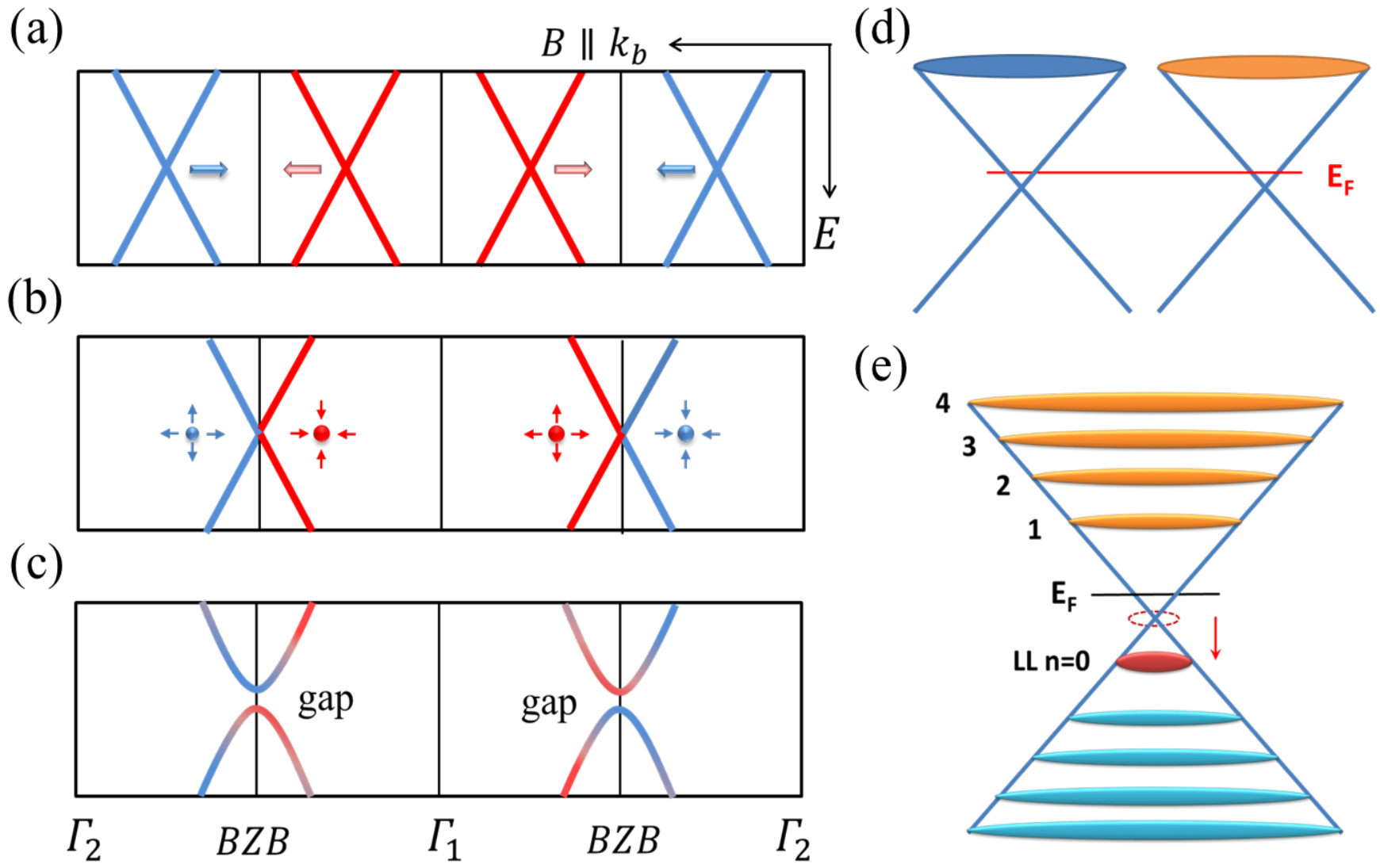

FIG. 3 (Color online). Zeeman splitting and Orbital effect. (a) Zeeman splitting will shift the two opposite Weyl points from Brillouin Zone center to Brillouin Zone boundary (BZB), where the two Weyl points with opposite chiralities eventually meet (b) and annihilate each other at the zone boundary and open a massive gap (c). $\Gamma_1$ and $\Gamma_2$ are the center of the first and second Brillouin Zone, respectively. (d) A schematic illustration of the Landau levels of two Weyl nodes without the orbital effect. (e) A schematic illustration of the orbital splitting of the massive 2D Dirac metal for zeroth Landau level. Due to the orbital effect, the zeroth LL shifts, for example, downward (depending on the chirality of the Dirac point), and introduces extra carriers.

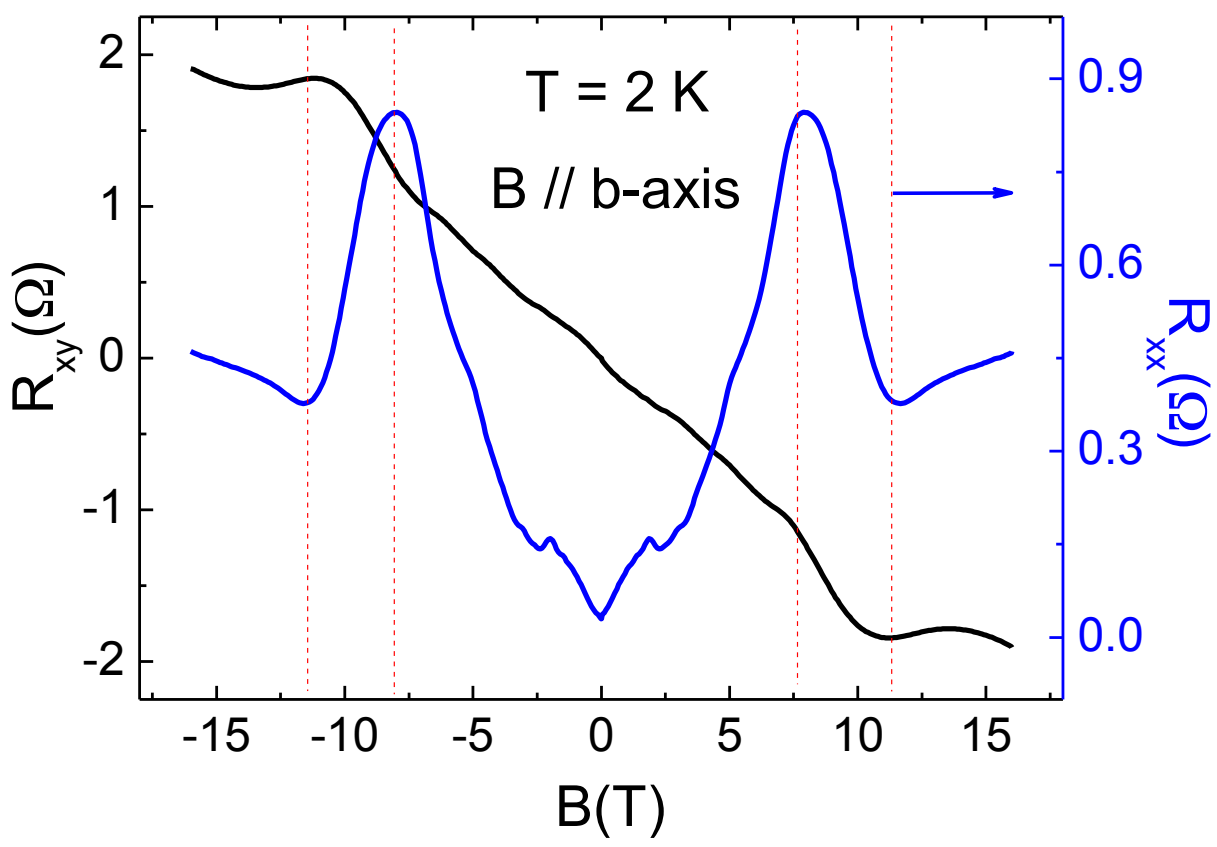

FIG. 4 (Color online). Hall effect in sample ♯2. Hall resistance exhibits an anomaly near the MR peak position, reveals that extra carriers are induced due to the orbital effect.

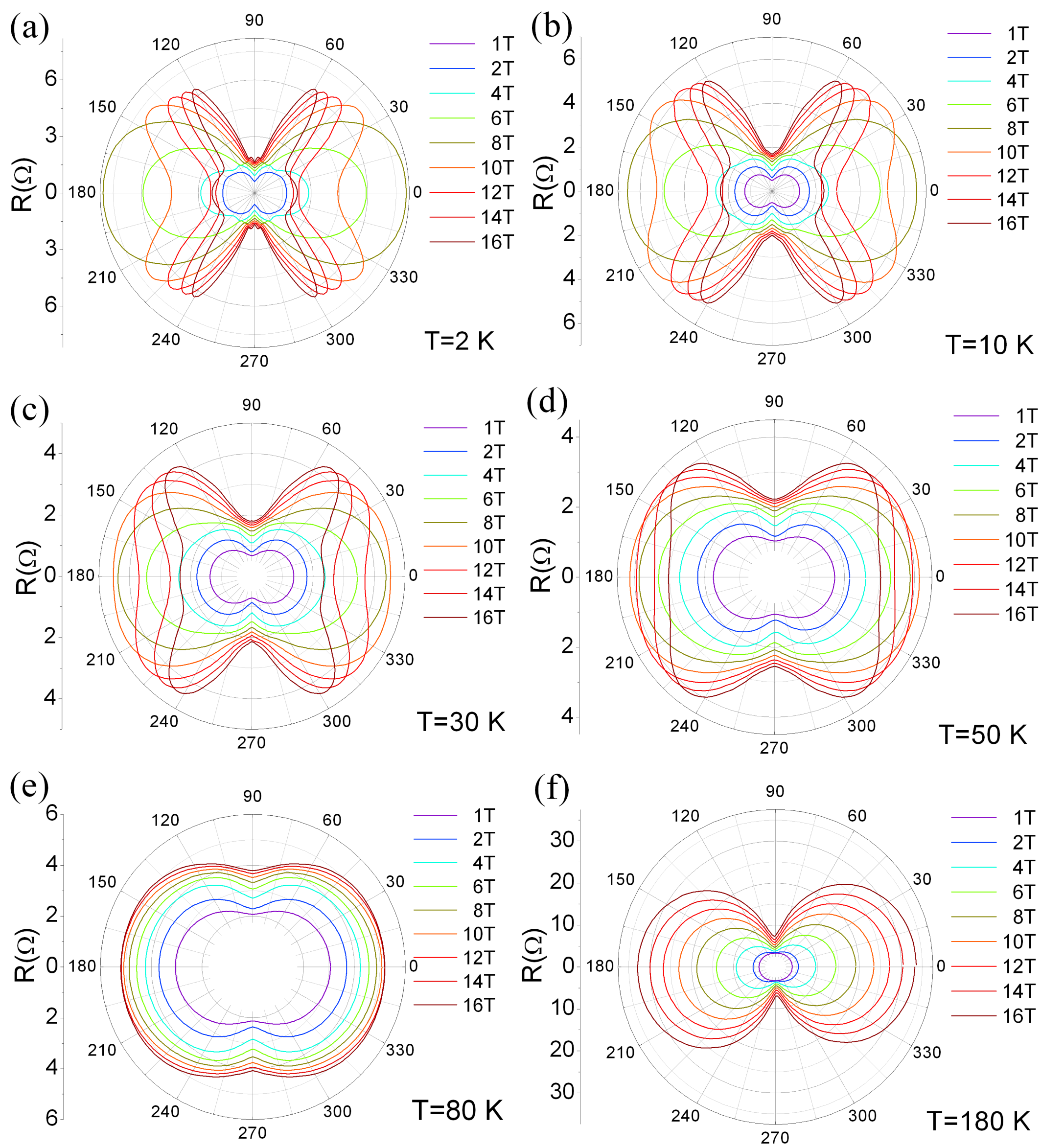

FIG. 5 (Color online). The AMR for sample ♯1 in *b*-*c* plane by rotating magnetic field under different temperatures. (a)-(f) are respectively obtained at 2, 10, 30, 50, 80 and 180 K.